\documentclass[twocolumn,showpacs,preprintnumbers]{revtex4}
\usepackage{amssymb}
\usepackage{amsfonts}
\usepackage{amsmath}
\usepackage{graphicx}
\usepackage{dcolumn}
\usepackage{bm}

\begin{document}

\title{Spin-splitting in the quantum Hall effect of disordered GaAs layers
with strong overlap of the spin subbands}
\author{S. S. Murzin$^{\text{1,2}}$, M. Weiss$^{\text{2}}$, D. A. Knyazev$^{%
\text{1}}, $ A. G. M. Jansen$^{\text{2,3}}$, and K.
Eberl$^{\text{4}}$} \affiliation{$^{\text{1}}$Institute of Solid
State Physics RAS, 142432, Chernogolovka,
Moscow District., Russia\\
$^{\text{2}}$Grenoble High Magnetic Field Laboratory, Max-Planck-Institut f%
\"{u}r Festk\"{o}rperforschung and Centre National de la Recherche
Scientifique, BP 166, F-38042, Grenoble Cedex 9, France \\
$^{\text{3}}$Service de Physique Statistique, Magn\'{e}tisme, et
Supraconductivit\'{e}, D\'{e}partement de Recherche \\
Fondamentale sur la Mati\`{e}re Condens\'{e}e, CEA-Grenoble, 38054 Grenoble
Cedex 9, France\\
$^{\text{4}}$Max-Planck-Institut f\"{u}r Festk\"{o}rperforschung, Postfach
800 665 D-70569, Stuttgart, Germany}

\begin{abstract}
With minima in the diagonal conductance $G_{xx}$ and in the absolute value
of the derivative $|dG_{xy}/dB|$ at the Hall conductance value $%
G_{xy}=e^{2}/h$, spin-splitting is observed in the quantum Hall effect of
heavily Si-doped GaAs layers with low electron mobility $\mu \approx 2000$ cm%
$^{2}$/Vs in spite of the fact that the spin-splitting is much smaller than
the level broadening. Experimental results can be explained in the frame of
the scaling theory of the quantum Hall effect, applied independently to each
of the two spin subbands.
\end{abstract}

\maketitle

In an electron system with a small $g-$factor strong disorder broadens and
suppresses the spin-splitted structure in the electron spectrum in an
applied magnetic field. Therefore, spin-splitting with Zeeman energy
separation $E_{s}=g\mu _{B}B$ ($\mu _{B}$ is the Bohr magneton and $B$ the
magnetic field) does not show up in the kinetic and thermodynamic properties
of strongly disordered 3D bulk electron systems. However, for a 2D system,
the scaling theory for diffusive interference effects leads to a quite
unexpected conclusion: the spin-splitting can arise in the magnetoquantum
transport data even in the case of very small $E_{s}$ with respect to
energy-level broadening $\Gamma $ if the temperature is sufficiently low.
For this situation with $E_{s}\ll \Gamma $, the spin-splitted quantum Hall
effect (QHE) with odd integer Hall-conductance plateaux at $%
G_{xy}=(2i+1)e^{2}/h$ and corresponding minima in the diagonal conductance
(per square) $G_{xx}$ should develop at low temperatures due to the
existence of extended states near the center of two spin-splitted Landau
levels with localized states in between \cite{Khm, Kag}. However, the
spin-splitting was not observed in disordered 2D GaAs systems with
mobilities below $10000$ cm$^{2}$/Vs \cite{ex} when $E_{s}\ll \Gamma $.
Higher mobility samples generally do show the spin-splitting \cite{Nich} due
to the enhanced Zeeman splitting in such pure two dimensional samples
because of electron-electron interaction effects \cite{Smith}. Strong
disorder should suppress this enhancement of spin-splitting \cite{Murthy}.

In the present work we observed for the first time the manifestation of the
spin-splitting in the magnetoconductance of a strongly disordered system, a
heavily Si-doped GaAs layer with a low electron mobility $\mu \approx 2000$%
~cm$^{2}$/Vs. For the case of these layers, the spin-splitting $E_{s}$, with
$E_{s}/k_{B}$ about $4$~K at a magnetic fields of $B=12\div 13$~T, is much
smaller than the level broadening $\Gamma $\ ($\gtrsim 100$~K) resulting in
a strong overlap of the two spin subbands. We analyzed the scaling
properties of the transport data of our electron system with weak
spin-splitting, assuming that the conductances of the different spin
subbands are renormalized independently for variations due to diffusive
interference effects. Such an approach is justified in the absence of
spin-flip scattering, at least, for non-interacting electrons. Experimental
data are in accordance with such an analysis.

\begin{figure}[t]
\includegraphics[width=8cm,clip]{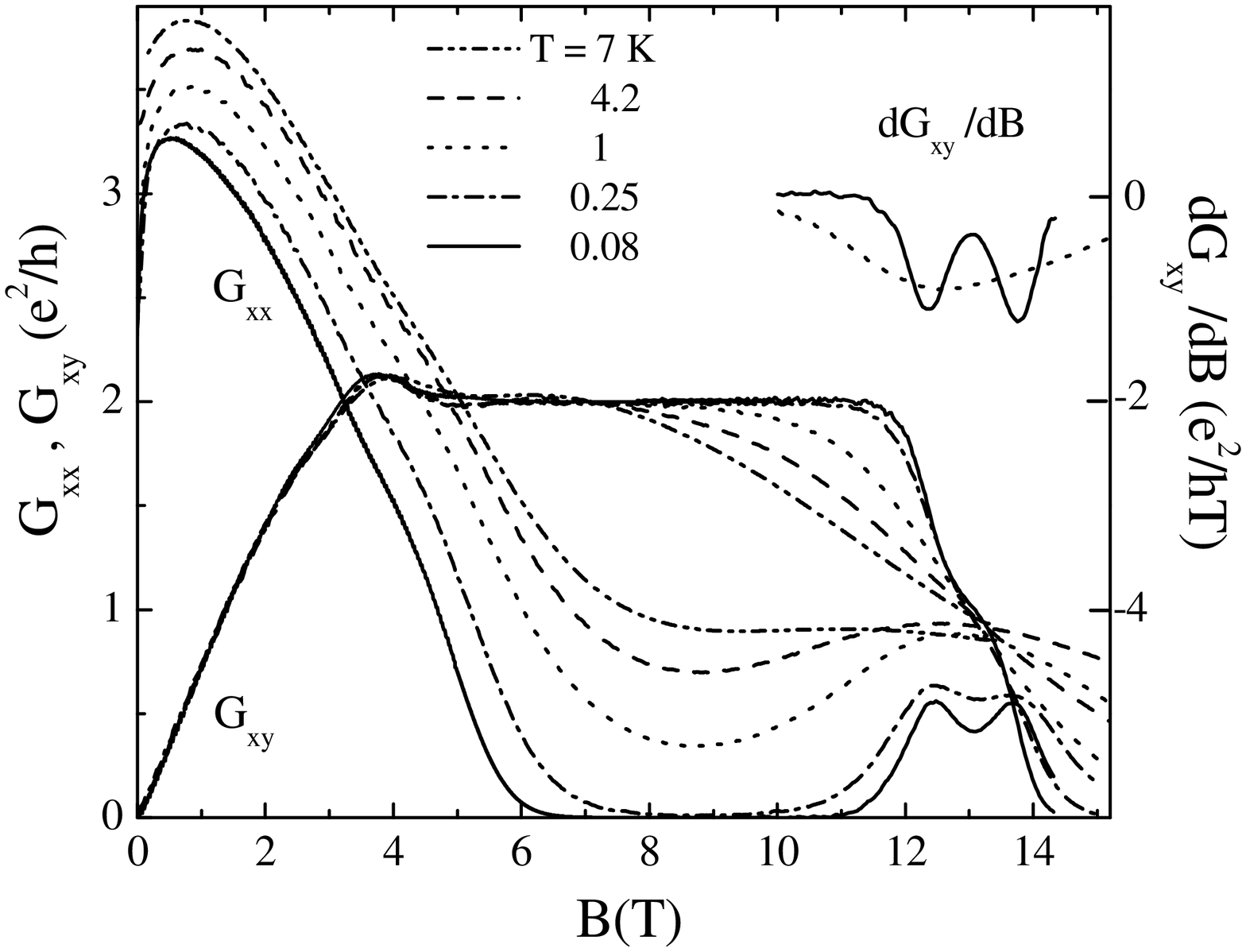}
\caption{Magnetic field dependence of the diagonal ($G_{xx}$) and Hall ($%
G_{xy}$) conductance, and of derivative $dG_{xy}/dB$ for sample~40
in a magnetic field perpendicular to the heavily doped GaAs layer
(thickness 40~nm) at different temperatures, showing
spin-splitting for $G_{xy}=1$ at 13~T.} \label{G40}
\end{figure}

The investigated heavily Si-doped n-type GaAs layers sandwiched between
undoped GaAs were prepared by molecular-beam epitaxy. The number given for a
sample corresponds to the thickness $d$ of the conducting doped layers with $%
d=34$, 40, and 50~nm. The Si-donor concentration is $1.5\times 10^{17}$ cm$%
^{-3}$. Hall bar geometries of width 0.2~mm and length 2.8~mm were etched
out of the wafers. A phase sensitive ac-technique was used for the
magnetotransport measurements down to 40~mK with the applied magnetic field
up to 20~T perpendicular to the layers.

\begin{figure}[t]
\includegraphics[width=6cm,clip]{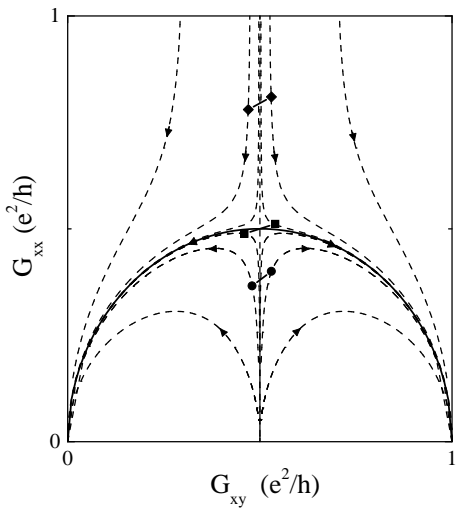}
\caption{Theoretical flow lines \protect\cite{Dolan} showing the coupled
evolution of the diagonal ($G_{xx})$ and Hall conductance ($G_{xy}$) for a
totally spin polarized electron system. The arrows indicate the direction of
flow for decreasing temperature. Pairs of diamonds, squares and circles
indicate possible starting positions ($G_{xy}^{-0},G_{xx}^{-0}$) and ($%
G_{xy}^{+0},G_{xx}^{+0}$) for the scaling in the cases (i), (ii), and (iii),
respectively, as described in the text. }
\label{flow}
\end{figure}

The electron densities per square as derived from the slope of the Hall
resistance $R_{xy}$ in weak magnetic fields ($0.5-3$~T) at $T=4.2$~K are $%
N_{s}=3.9$, 4.6 and $5.0\times 10^{11}$ cm$^{-2}$ for samples 34,
40 and 50, respectively. The bare\ high temperature mobilities
$\mu _{0}$ are about 2000, 2200 and $2400$~cm$^{2}$/Vs. Because of
the rather large quantum corrections to the conductance, even in
zero magnetic field at 4.2 K, we used for determining the mobility
the approximate relation $\mu _{0}=R_{xy}/BR_{xx}$ at the
intersection point of the $R_{xx}(B)$ curves for different
temperatures.

The characteristic energy scales of our samples with not more than
two size-quantized energy levels are as follows. The Fermi energy
at zero magnetic field $E_{F}/k_{B}\approx 200$~K, the splitting
of the size quantization $E_{sq}/k_{B}=3(\pi \hbar
/d)^{2}/2mk_{B}\approx 100\div 200$~K (for our thinnest sample
with $E_{sq}/k_{B}\approx 200$~K the second subband is occupied
due to disorder), $\hbar /\tau k_{B}\approx 100$~K ($\tau $ is the
transport relaxation time at zero magnetic field), the
Landau-level
energy broadening $\Gamma /k_{B}=\hbar \sqrt{2\omega _{c}/\pi \tau }%
/k_{B}\approx 130$~K, and cyclotron energy $\hbar \omega _{c}/k_{B}\approx
250$~K at the magnetic field $B=12\div 13$~T.

The magnetoconductance data for the three samples are rather similar. In Fig.%
\ref{G40} the diagonal ($G_{xx}$) and Hall ($G_{xy}$) conductance as
calculated from the diagonal per square ($R_{xx}$) and Hall ($R_{xy}$)
resistance have been plotted for sample~40. At low temperatures the curves $%
G_{xy}(B)$ show a wide QHE plateau from $\approx 6$ up to $\approx 11.5$~T
with the value of $G_{xy}=2$ accompanied by an exponentially small value of $%
G_{xx}$ at low temperatures $T\lesssim 0.3$ K \cite{M01} . At the lowest
temperature the diagonal ($G_{xx}$) conductance and derivative $|\partial
G_{xy}/\partial B|$ show minima at $B\approx 13$ T where at high temperature
the Hall conductance $G_{xy}\approx 1$. The value of the magnetic field
where the spin-splitted QHE is observed is only $1.5$ times larger than the
QHE field with $G_{xy}=2$, which should be compared with the factor 2
expected from filling-factor related arguments.

\begin{figure}[t]
\includegraphics[width=8.5cm,clip]{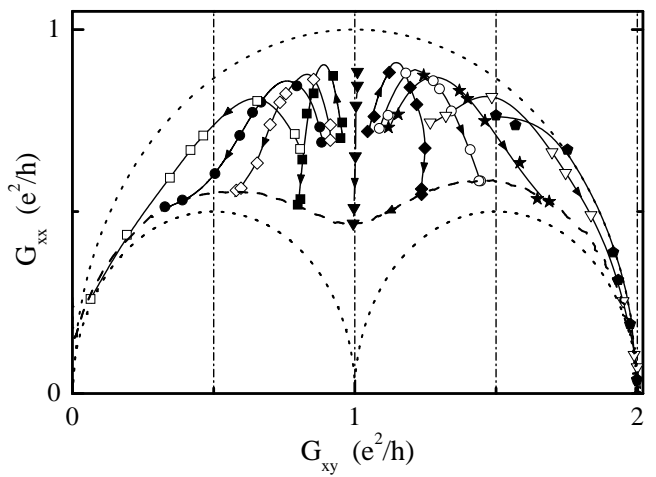}
\caption{Flow-diagram of the ($G_{xx}(T),G_{xy}(T)$) data points
for sample 34 with decreasing temperature (arrows) from $12$ down
to $0.1$ K. Different symbols connected by solid lines are for
different magnetic fields from 9 to 13~T. Dotted lines show the
semicircles (Eq.\protect\ref{sl1}) and a two times larger one. The
dashed line shows the magnetic-field-driven diagonal conductance
$G_{xx}(G_{xy})$ from 8 to 13.8~T at $T=0.1$ K. Vertical
dash-dotted grid lines are plotted for reference.} \label{fl34}
\end{figure}

The scaling treatment of the QHE \cite{Pr} results in a graphical
presentation of the flow diagram \cite{Khm1} which depicts the coupled
evolution of the diagonal ($G_{xx}$) and Hall ($G_{xy}$) conductance
components with increasing coherence length. Recent developments \cite{Dolan}
of the scaling theory based on symmetry arguments resulted in a calculation
of the exact shape of the flow lines $G_{xx}(G_{xy})$ for a totally spin
polarized electron system as plotted in Fig.\ref{flow} for $0\leq G_{xy}\leq
1$. The different quantum Hall phases ($i=0,1...$) in the flow diagram are
separated by the vertical lines $G_{xy}=i+1/2$. At sufficiently low
temperatures the $(G_{xx},G_{xy})$ data flow on a separatrix in the form of
a semicircle
\begin{equation}
G_{xx}^{2}+\left[ G_{xy}-(i+1/2)\right] ^{2}=1/4.  \label{sl1}
\end{equation}%
Critical points can be found at $(G_{xy}^{c},G_{xx}^{c})=(i+1/2,1/2)$. The
same critical positions were found in microscopic descriptions of the QHE
for the case of non-interacting electrons \cite{Huo,RF}.

In Fig.\ref{fl34} we have plotted the flow lines showing the temperature
evolution of the points $(G_{xy}(T),G_{xx}(T))$ of conductance for sample 34
at different magnetic fields with temperature ranging from $\approx 10$ down
to $\approx 0.1$~K. For sample 40, the flow diagram is rather similar to one
for sample 34. For these samples, at the magnetic fields where the
spin-splitting is observed, the flow lines move upwards and than downwards
for decreasing temperatures. The lines cross each other for data at
different magnetic fields, in contrast to the theoretical prediction for the
case of a totally spin polarized electron system (see Fig.\ref{flow}). For
sample 50, the flow lines do not show the upward trend and are not crossing
each other. For low temperatures (below $3$ K) the flow diagrams are very
similar for all three samples: the flow lines approach the semicircles
according Eq.\ref{sl1}. Linear extrapolation of $G_{xx}(T)$ and $G_{xy}(T)$
from 0.5 to $0$~K at the two fields where $%
G_{xx}(B)$ has a maximum (see Fig.\ref{G40}) results in values $%
G_{xx}=0.5\pm 0.02$, $G_{xy}=0.5\pm 0.05$ and $1.5\pm 0.05$. These critical
values are the same as predicted for a totally spin polarized electron
system. At the lowest temperatures, the magnetic-field driven dependence $%
G_{xx}(G_{xy})$ is mostly not far from the two smaller semicircles (Eq.\ref%
{sl1}) shown in Fig.\ref{fl34} by the dashed lines.

In the absence of spin-flip scattering, the conductances of the different
spin subbands are renormalized independently, at least, for the case of
non-interacting electrons. Since the temperature dependence of the
magnetoconductance is not known for a single spin-polarized band, it is
impossible to estimate accurately the flow lines for the total conductance
from the flow lines for the single polarized bands because the summation $%
G_{ij}(T)=G_{ij}^{-}(T)+G_{ij}^{+}(T)$ involves different positions on the
spin-polarized flow lines at the same temperature. The index $+$ and $-$\
correspond to the majority and minority spin subsystems with larger and
smaller Hall conductances, respectively. Nevertheless we can do some
conclusions about the scaling properties of the total conductance $G_{ij}$.

For weak spin-splitting $g\mu _{B}B\ll \hbar /\tau \lesssim E_{F}$ the bare
(non-renormalized) conductances $G_{ij}^{0\pm }$ for the two spin subbands
as measured at high temperatures
\begin{equation}
G_{ij}^{0\pm }=\frac{G_{ij}^{0}}{2}\pm \frac{g\mu _{B}B}{4}\frac{\partial
G_{ij}^{0}}{\partial E},
\end{equation}%
differ weakly from each other because $g\mu _{B}B\partial
G_{ij}^{0}/\partial E\sim G_{ij}^{0}g\mu _{B}B/\Gamma \ll G_{ij}^{0}$. Here $%
G_{ij}^{0}=G_{ij}^{0-}+G_{ij}^{0+}$ and $E_{F}$ the Fermi energy. The QHE
with total Hall conductance $G_{xy}=1$ should arise when one subsystem is in
the insulator state ($(G_{xy}^{-},G_{xx}^{-})\rightarrow (0,0)$ for $%
T\rightarrow 0$) (see Fig.\ref{flow} for the conductances of a single
spin-polarized band), and the other in the QHE state ($%
(G_{xy}^{+},G_{xx}^{+})\rightarrow (1,0)$). This occurs in a narrow magnetic
field range where $G_{xy}^{+0}>1/2$ but $G_{xy}^{-0}<1/2$. At the critical
value of $G_{xy}^{-0}=1/2$ and $G_{xy}^{+0}>1/2$, $(G_{xy}^{-},G_{xx}^{-})%
\rightarrow (1/2,1/2)$ and $(G_{xy}^{+},G_{xx}^{+})\rightarrow (1,0)$,
therefore, the total conductance $(G_{xy},G_{xx})\rightarrow (3/2,1/2)$.
Similarly, at the critical value of $G_{xy}^{+0}=1/2$, the total conductance
$(G_{xy},G_{xx})\rightarrow (1/2,1/2)$. Thus, the critical points are the
same as for the case of a totally spin polarized electron system, in
accordance with experimental results. This differs from the situation for
the distorted flow diagram predicted for the case of small spin-splitting
compared to the cyclotron energy ($g\mu _{b}B<\hbar \omega _{c}$ \cite{Ds}),
leading to essentially different positions of the critical points who's
exact position depends on the amount of spin-splitting. Note, that these
results \cite{Ds} have been obtained on the basis of a postulated symmetry
group in order to include spin-splitting, without giving any microscopic
picture for the scaling behavior.

\begin{figure}[t]
\includegraphics[width=7cm,clip]{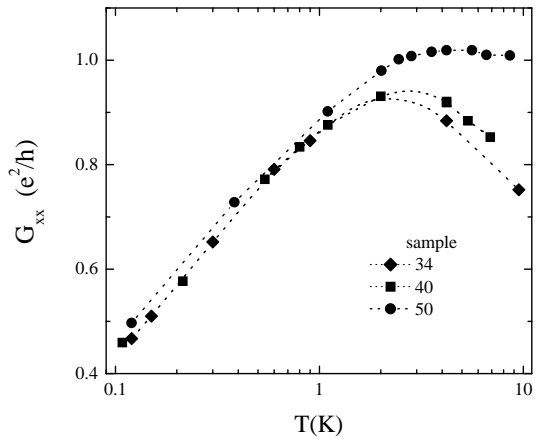}
\caption{Temperature dependence of the diagonal ($G_{xx}$) conductance, for
sample~34 (diamonds), 40 (squares), and 50 (circles) at a magnetic field $%
B=13.1$~T where the spin-splitted structure is observed.}
\label{temp}
\end{figure}

At low enough temperatures, when the spin-splitted QHE is well developed so
that in the QHE minimum $G_{xx}\approx 0$, one can argue that the flow lines
should follow the lines derived for the case of a totally spin polarized
electron system. In the minimum of $G_{xx}$ holds $%
(G_{xy}^{-},G_{xx}^{-})=(0,0)$ and $(G_{xy}^{+},G_{xx}^{+})=(1,0)$, i.e. the
minority subsystem does not contribute to conductance and the majority
subsystem contributes only the quantum value $G_{xy}=1$ to the Hall
conductance. At lower magnetic fields, the $+$ subsystem contributes only to
the Hall conductance the value $1$ as before, and the total conductance $%
(G_{xy},G_{xx})=(G_{xy}^{-}+1,G_{xx}^{-})$. Similarly, at higher magnetic
fields $(G_{xy},G_{xx})=(G_{xy}^{+},G_{xx}^{+})$. At the lowest
temperatures, the total conductance ($G_{xy}$,$G_{xx}$) is expected to flow
along the same lines as derived for a single spin-polarized electron system.
Therefore, $G_{xx}$ as a function of $G_{xy}$ flows for a changing magnetic
field close to the semicircles given by Eq.\ref{sl1}, in accordance with
experimental data below 0.1~K.

As mentioned above, for the totally spin polarized electron system, the flow
lines should not cross each other \cite{Khm1} in contrast to our
experimental data. For the case of two different spin projections, we will
show now that the flow lines starting in the region in between the large
semicircle and the two smaller semicircles shown in Fig.\ref{fl34} can cross
each other. Consider the line ending at the critical point $(0.5,0.5)$ as a
reference line. Starting slightly at the left from the starting point of
this line, and knowing that this line should end at (0,0), the crossing is
unavoidable. Starting at the right from the starting point of this reference
line, and knowing that this line should end at $(1,0)$, leads also to a
crossing point.

The scaling theory predicts different type of the temperature dependence $%
G_{xx}(T)$ in the magnetic-field region where $G_{xy}^{0}$ is close to $0.5$
for the case of the totally spin-polarized electron system. (i) For $%
G_{xx}^{0}>0.5$ flow lines go down , i.e. $G_{xx}$ decreases with
decreasing temperature as shown in Fig.\ref{flow} for points
starting at the diamonds. (ii) For $G_{xx}^{0}\approx 0.5$ flow
lines start nearly horizontally near the maximum of the semicircle
(starting at the squares in Fig.\ref{flow}) leading to a very weak
temperature dependence of $G_{xx}(T)$ followed by a decrease at
lower temperatures. (iii) For $G_{xx}^{0}<0.5$ the
flow lines go up to a value of $G_{xx}$\ close to $0.5$ and than down, i.e. $%
G_{xx}$ at first increases with decreasing temperature and than decreases
with a maximum value of the diagonal conductance $\approx 0.5$. For the case
of two spin-subbands with small spin-splitting the above description remains
valid for the conductances of the two spin polarizations adding up to the
measured total conductance. The Hall conductance should depend weakly on
temperature in all these cases, because $G_{xy}^{-}$ decreases and $%
G_{xy}^{+}$ increases with decreasing temperature.

In Fig.\ref{temp} we plot the diagonal conductance $G_{xx}$ as a function of
temperature $T$\ in the spin-splitted minima of $G_{xx}$ where $G_{xy}^{0}$ $%
\approx 1$ for the three samples. The data are in accordance with the above
given prediction. For samples 34 and 40, $G_{xx}$ is a non-monotonic
function of temperature with a maximum value of $G_{xx}$ slightly lower than
$1$ around 3~K, corresponding to the case (iii). For sample 50, $%
G_{xx}\approx 1$ does not change in the high-temperature range above 3~K and
decreases at lower temperatures, corresponding to the case (ii). Note, the
temperature dependences of $G_{xx}$ are different for layers, which differ
from each other only by thickness.

At magnetic fields near the spin-splitted QHE structure, the localization
lengths $\xi ^{\pm }$ of the two spin systems are large because both spin
systems are close to the quantum Hall state-insulator transition ($%
|1/2-G_{xy}^{0\pm }|\ll 1$), where $\xi ^{\pm }\rightarrow \infty $.
Therefore, for the observation of the QHE with $G_{xy}=1$ a much lower
temperature (or a larger coherence length) is necessary than for the QHE
with $G_{xy}=2$. In previous experiments at smaller magnetic fields \cite{ex}
the spin-splitting was not observed, probably, because at smaller fields $%
\mu _{B}gB\partial G_{ij}^{0}/\partial E$ is smaller. In our samples the
spin-splitting is observed only at low temperatures $T\lesssim 0.1$ K, what
is\ much smaller than $\hbar /\tau \approx 100$ K, and even than $%
E_{s}/k_{B}\approx 4$~K.

The assumption about independent renormalization of the
conductances of the two spin subbands is undoubtedly valid for
non-interacting electrons in the absence of spin-flip scattering.
Although electron-electron interaction is important in real
systems, the experimental study of the flow diagram on samples 34,
40, and other thinner layers \cite{fl} shows good quantitative
agreement with the predicted flow lines \cite{Dolan} for half the
measured conductance values in the field range below 6~T, where
there is not any manifestation of spin-splitting and, therefore,
$G_{ij}/2=G_{ij}^{+}=G_{ij}^{-}$. This gives support for our model
of independent spin-band contributions leading to the same
critical points as for a spin-polarized system.

In summary, we observed spin-splitting in the QHE of heavily doped n-type
GaAs layers with disorder much larger than spin-splitting $g\mu _{B}B$. Our
results are in accordance with the scaling treatment of the quantum Hall
effect, applied independently to the two spin subbands. Namely, the magnetic
field position for the QHE is imposed by the occurrence of the Hall quantum
value $G_{xy}\approx 1$, although this field position is only $1.5$ times
larger than the QHE field with $G_{xy}=2$. Several features in the ($G_{xy}$,%
$G_{xx}$) flow diagrams, like the observed critical values $%
G_{xx}^{c}=0.5\pm 0.02$, $G_{xy}^{c}=0.5\pm 0.05$\ and $1.5\pm 0.05$ and the
anomalous shapes of the flow lines, can be deduced from an independent
summation of the contributions of the two spin bands. The spin-splitting is
observed at temperatures $T\lesssim 0.1$ K much smaller than all other
energy scales determining the electron spectrum.

This work is supported by the Russian Foundation for Basic Research. We
would like to thank B. Lemke for her help in the preparation of the samples.


\begin{thebibliography}{99}
\bibitem{Khm} D. E. Khmel'nitski\u{\i}, Helvetica Phys. Acta \textbf{65},
164 (1992).

\bibitem{Kag} V. Kagalovsky, B. Horovitz, and Y. Avishai, Phys. Rev. B
\textbf{55}, 7761 (1994), and references therein.


\bibitem{ex} H. W. Jiang, C. E. Johnson, K. L. Wang, and S. T. Hannahs,
Phys. Rev. Lett. \textbf{71}, 1439 (1993); T. Wang, K. P. Clark, G. F.
Spencer, A. M. Mack, and W. P. Kirk, ibid. \textbf{72}, 709 (1994); J. F.
Hughes, J. T. Nicholls, J. E. F. Frost, E. H. Linfield, M. Pepper, C. J. B.
Ford, D. A. Ritchie, G. A. C. Jones, E. Kogan, and M. Kaveh, J. Phys.
Condens. Matter \textbf{6}, 4763; (1994); C. H. Lee, Y. H. Chang, Y. W.
Suen, and H. H. Lin, Phys. Rev. B \textbf{58}, 10 629 (1998); C. F. Huang,
Y. H. Chang, C. H. Lee, H. T. Chou, H. D. Yeh, C.-T. Liang, Y. F. Chen, H.
H. Lin, H. H. Cheng, and G. J. Hwang, Phys. Rev. B \textbf{65}, 045303
(2001).

\bibitem{Nich} R. J. Nicholas et al., Phys. Rev. B \textbf{37}, 1294 (1988),
and references therein.

\bibitem{Smith} A. P. Smith, A. H. MacDonald, and G. Gumbs, Phys. Rev. B
\textbf{45}, 8829 (1992), and references therein.

\bibitem{Murthy} G. Murthy, Phys. Rev. B \textbf{64}, 241309 (2001).

\bibitem{Dolan} B. P. Dolan, Nucl. Phys. B \textbf{554}, \emph{487} (1999);
cond-mat/9809294.

\bibitem{M01} S. S. Murzin, M. Weiss, A. G. M. Jansen, and K. Eberl, Phys.
Rev. B \textbf{64}, 233309 (2001).

\bibitem{Pr} A. M. M. Pruisken, in \emph{The Quantum Hall Effect}, edited by
R. E. Prange and S. M. Girven, Springer-Verlag, 1990.

\bibitem{Khm1} D. E. Khmel'nitskii, Pis'ma Zh. Eksp. Teor. Fiz. \textbf{38},
454 (1983), [JETP Lett. \textbf{38}, 552 (1984)]; Phys. Lett. A \textbf{106}%
, 182 (1984).

\bibitem{Ds} B. P. Dolan, Phys. Rev. B \textbf{62}, 10278 (2000).

\bibitem{Huo} Y. Huo, R. E. Hetzel, and R. N. Bhatt, Phys.\ Rev.\ Lett.\
\textbf{70}, 481 (1993).

\bibitem{RF} Igor Ruzin and Shechao Feng, \ Rev.\ Lett.\ \textbf{74}, 154
(1995).

\bibitem{fl} S. S. Murzin, M. Weiss, A. G. M. Jansen and K. Eberl, Phys.
Rev. B \textbf{66}, 233314 (2002).
\end{thebibliography}
\end{document}